\documentclass[letterpaper,twocolumn,10pt]{article}
\usepackage{usenix2019_v3}

\usepackage{tikz}
\usepackage{amsmath}
\usepackage{graphics,hyphenat}
\usepackage{ifthen}
\usepackage{url}
\usepackage{enumitem}
\usepackage{color}
\usepackage{etoolbox}
\usepackage{xspace}
\usepackage{makecell}
\usepackage{subcaption}
\usepackage{amssymb} % checkmark
\usepackage{fancyhdr}
\usepackage[stable]{footmisc}

\begin{document}
\newcommand{\itempar}[1]{ \noindent \textbf{#1}}
\newcommand{\mysec}[1]{\section{#1}\vspace{0in}}
\newcommand{\mypar}[1]{\noindent \textbf{#1:}}
\newcommand{\myitem}{\vspace{-0.1cm} \item}
\newcommand{\note}[1]{\textbf{NOTE: [#1]}}
\newcommand{\NOTE}[1]{\textbf{NOTE: [#1]}}
\newcommand{\todo}[1]{\textbf{TODO: [#1]}}
\newcommand{\TODO}[1]{\textbf{TODO: [#1]}}
\newcommand{\fixme}[1]{\textbf{FIXME: [#1]}}
\newcommand{\Fixme}{{\bf FIXME:}}
\newcommand{\remark}[1]{\textbf{REMARK [#1]???}} 
\newcommand{\response}[1]{\textbf{RESPONSE [#1]???}} 
\newcommand{\dropme}[1]{\textbf{DROPME [#1]???}} 
\newcommand{\reviewcomment}[2]{\textbf{REVIEW \##1}{\xspace\emph{#2}}} %\newcommand{\WORDS}[1]{#1}
\newcommand{\WORDS}[1]{}
\newcommand{\WORDSresponse}[1]{#1}
\newcommand{\WORDSchange}[1]{#1}

\newcommand{\eg}{\textit{e.g.}}
\newcommand{\ie}{\textit{i.e.}}
\newcommand{\etal}{\textit{et al.}\xspace}
\newcommand{\codesm}[1]{\texttt{\small #1}}
\newcommand{\ignore}[1]{}
\newcommand{\budget}[1]{\textbf{SPACE BUDGET: #1}}
\newcommand{\myincludegraphics}[2]{\resizebox{#1}{!}{\includegraphics{#2}}}
\newcommand{\red}[1]{{\color{red} #1}}
\newcommand{\haris}[1]{{\color{red} #1}}

% convenient macros for use with algorithmic
\newcommand*\BitAnd{\mathrel{\&}}
\newcommand*\BitOr{\mathrel{|}}
\newcommand*\ShiftLeft{\ll}
\newcommand*\ShiftRight{\gg}
\newcommand*\BitNeg{\ensuremath{\mathord{\sim}}}

% convenient macros for 
\newcommand{\clwb}{\texttt{clwb}\xspace}
\newcommand{\clflush}{\texttt{clflush}\xspace}
\newcommand{\clflushopt}{\texttt{clflushopt}\xspace}
\newcommand{\pcommit}{\texttt{pcommit}\xspace}
\newcommand{\epoch}{\texttt{epoch}\xspace}
\newcommand{\fence}{\texttt{fence}\xspace}
\newcommand{\mov}{\texttt{mov}\xspace}
\newcommand{\movnti}{\texttt{movnti}\xspace}
\newcommand{\sfence}{\texttt{sfence}\xspace}
\newcommand{\lazypcommit}{\texttt{lazy\_pcommit}\xspace}
\newcommand{\cas}{\texttt{cas}\xspace}
\newcommand{\bench}{PMBench\xspace}
\newcommand{\SelfService}{Memory-oriented distributed computing\xspace}
\newcommand{\Mediated}{Message-based distributed computing\xspace}
\newcommand{\selfService}{memory-oriented distributed computing\xspace}
\newcommand{\mediated}{message-based distributed computing\xspace}
\newcommand{\MODC}{MODC\xspace}
\newcommand{\shelf}{region\xspace}
\newcommand{\shelves}{regions\xspace}
\newcommand{\copyonwrite}{copy-on-write\xspace}

\newboolean{expandConfidential}
\setboolean{expandConfidential}{true}
%\setboolean{expandConfidential}{false}

\ifthenelse {\boolean{expandConfidential}} 
{
\newcommand{\confidential}[1]{#1}
}
{
\newcommand{\confidential}[1]{}
}

% macros
\newcommand{\dm}{\douppercase{d}isaggregated memory\xspace}
\newcommand\douppercase[1]{\ifnum\ifhmode\spacefactor\else2000\fi>1000 \uppercase{#1}\else#1\fi}
\newcommand{\Dm}{Disaggregated memory\xspace}

\title{\vspace{-0.3in}\MODC: Resilience for disaggregated memory architectures \\using task-based programming}
\author{Kimberly Keeton$^1$, Sharad Singhal, Haris Volos$^2$, Yupu Zhang$^3$,\\
Ramesh Chandra Chaurasiya, Clarete Riana Crasta, Sherin T George,\\ 
Nagaraju K N, Mashood Abdulla K, Kavitha Natarajan, Porno Shome, Sanish Suresh\\
\small {\em  Hewlett Packard Enterprise} \\
}
\date{}
\maketitle

% A hack to construct footnotes without marker taken from here: 
% https://tex.stackexchange.com/questions/30720/footnote-without-a-marker
\newcommand\blfootnote[1]{%
  \begingroup
  \renewcommand\thefootnote{}\footnote{#1}%
  \addtocounter{footnote}{-1}%
  \endgroup
}

\blfootnote{\hspace{-0.5cm}Authors' current affiliations:}
\blfootnote{\hspace{-0.5cm}$^1$Unaffiliated, $^2$University of Cyprus, $^3$Google}
% $^4$Nvidia}

% Need to reclaim some vertical space because the above footnotes push down the Abstract 
\vspace{-0.6cm}
\begin{abstract}
{
Disaggregated memory architectures provide benefits to applications beyond traditional scale out environments, such as independent scaling of compute and memory resources. 
They also provide an independent failure model, where computations or the compute nodes they run on may fail independently of the disaggregated memory; thus, data that’s resident in the disaggregated memory is unaffected by the compute failure. 
Blind application of traditional techniques for resilience (e.g., checkpoints or data replication) does not take advantage of these architectures. To demonstrate the potential benefit of these architectures for resilience, we develop \emph{Memory-Oriented Distributed Computing (\MODC)}, a framework for programming disaggregated architectures that borrows and adapts ideas from task-based programming models, concurrent programming techniques, and lock-free data structures.
This framework includes a task-based application programming model and a runtime system that provides scheduling, coordination, and fault tolerance mechanisms. 
We present highlights of our \MODC prototype and experimental results demonstrating that \MODC-style resilience outperforms \ignore {traditional resilience approaches} a checkpoint-based approach in the face of failures.
}
\end{abstract}
\mysec{Introduction}

\sloppy
Recent technology advances in high-density, byte-addressable non-volatile memory (NVM) (\eg,~\cite{3d-xpoint:intel:2015, strukov:memristor:nature:2008, qureshi:pcm:isca:2009, xie:nvm-survey:des-test-comp:2011, hpe:nv-dimm})
and low-latency interconnects (\eg,~\cite{rdma-consortium, genz-consortium, knebel:genz:hotchips:2019})
have enabled building rack-scale systems with a large disaggregated memory pool shared across decentralized compute nodes. 

Disaggregated memory architectures present independent failure domains and a partial failure model: computations or the compute nodes they run on may die, but disaggregated memory remains available. 
When a compute node fails, updates propagated to disaggregated memory remain visible to other compute nodes. 
This partial failure model avoids a complete system shutdown in the event of a component failure.
This failure model is an improvement over the failure model of scale up architectures, where a compute failure causes the failure of the entire application (and potentially the failure of the entire machine).
It's also an improvement over the failure model of scale out architectures, where a compute node failure may render a part of the dataset unavailable, unless the data is stored redundantly across nodes.

The applications that thrive in a disaggregated memory environment have large and long-lived datasets that aren't amenable to traditional resilience techniques such as checkpointing.
For example, large-scale graph analytics applications at social media companies like Facebook operate over graphs with billions of vertices and trillions of edges that consume hundreds of petabytes~\cite{bronson:tao:usenix:2013}.
Such applications may benefit from maintaining data in disaggregated memory for resilient computation on the graph data.

We develop \emph{Memory-Oriented Distributed Computing (\MODC)}, a framework for programming disaggregated architectures, by borrowing and adapting ideas from task-based programming models, concurrent programming techniques, and lock-free data structures. 
Task-based programming achieves fine-grained recoverability without explicit checkpoints. 
Developers write applications as a collection of idempotent tasks that operate on 
data items resident in disaggregated memory, and specify data and control dependencies between the tasks.
The underlying runtime system uses this dependency information to schedule the tasks such that dependencies are satisfied,
and to restart tasks if they fail.

The programming model is layered on top of a runtime system designed for a disaggregated environment where the disaggregated memory may not have general-purpose computational capabilities. 
As a result, we do not use server processes to actively manage 
scheduling, failure detection and coordination.
Instead, our approach 
leverages shared data structures in disaggregated memory to encapsulate global state. 
Workers in a user application access these concurrent data structures in a decentralized fashion through a library interface to the \MODC runtime. 
The runtime library uses one-sided operations and lock-free programming techniques to operate on the shared data structures, which ensures deadlock freedom and progress of the runtime in the event of compute failures. In particular, disaggregated memory holds dependency-tracking metadata and per-worker work queues, which enables work stealing to respond to failures and load imbalances.

We demonstrate that \MODC enables resilient applications with lightweight failure recovery (\eg, with less than 1\% penalty, as compared to failure-free execution). A checkpoint-based failure recovery approach is as much as 51\% slower than the \MODC approach.

\section{Background}

We begin by providing background on task-parallel programming models and disaggregated memory systems, as well as stating our system model.

\subsection{Task-parallel programming}
In task-parallel programming~\cite{blumofe:cilk:ppopp:1995}, the programmer decomposes an application into units of work, called \emph{tasks}, that may execute in parallel. 
Programmers focus on structuring their program to expose parallelism, while an underlying runtime system takes on the responsibility of dynamically scheduling and executing tasks among a pool of workers. 

\WORDSresponse{Traditionally, tasking systems have primarily relied on global checkpoint/restart mechanisms for tolerating task failures.
However, recent work has investigated other approaches to handling task failures}. 
Resilient X10~\cite{cunningham:resilient-x10:ppopp:2014} proposes extensions to the X10 task-parallel language~\cite{charles:x10:oopsla:2005} to expose failures to programmers, who can then handle individual task failures by exploiting domain-specific knowledge.
CIEL~\cite{murray:ciel:nsdi:2011} and Ray~\cite{moritz:ray:osdi:2018}, like us, provide transparent lineage-based fault tolerance. 
They log function calls and require arguments and results of tasks to be immutable, so that they can \WORDSresponse{use the lineage to} transparently restart individual tasks on the event of a failure.
A key difference is that our work leverages
disaggregated memory to fail over computation from a failed worker to another live worker in a decentralized manner, by letting live workers steal work from a failed worker.

\WORDSresponse{
Dataflow systems, such as MapReduce~\cite{dean:mapreduce:osdi:2004}, Dryad~\cite{isard:dryad:eurosys:2007}, and Spark~\cite{zaharia:spark:nsdi:2012} adopt a bulk-synchronous parallel (BSP) execution model, where all tasks within the same stage perform the same computation and cannot recursively create other tasks.
MapReduce limits the dataflow to a bipartite graph comprising map and reduce tasks. 
Dryad extends the MapReduce model to allow computation to be expressed into a more general directed acyclic graph (DAG), but lacks support for dynamic task graphs, which can be useful for iterative and incremental computation.
Spark does support dynamic task graphs, by dynamically tracking task dependencies through a centralized scheduler and adopting lineage-based fault tolerance. However, invoking the centralized scheduler at the end of each task adds overhead and increases latency. 
Streaming dataflow systems, such as Naiad~\cite{murray:naiad:sosp:2013} and Flink~\cite{carbone:flink:vldb:2017}, implement the dataflow using a graph of continuous operators or long running tasks, avoiding frequently invoking a centralized scheduler, but limiting support to static task graphs. 
}

\WORDSresponse{
\WORDSchange{Architectures based on dataflow languages~\cite{Dataflow} control program execution using \emph{actors} that consume \emph{objects} as inputs and produce objects as outputs.}
While these architectures also track dependencies among actors as part of program execution, 
the underlying actors are much more granular than our tasks, rely on specialized hardware for program execution and dependency management, and are not focused on resilience.
}

\subsection{Rack-scale disaggregated memory}

Resource disaggregation is a popular approach 
for specializing resources allocated to a workload. Increasing attention is being paid to leveraging memory resources outside  
of traditional compute nodes. Here we focus on \emph{disaggregated memory}, rather than remote memory solutions where compute nodes share their memory with other compute nodes.
In disaggregated memory system organizations, memory that is distinct from compute node memory is exposed over a fast network\ignore {, and can be scaled independently of the compute or storage capacity of the system}.
Disaggregated memory may be constructed with intelligent memory servers containing general-purpose processors (\eg,\cite{lim:memory-blade:isca:2009, lim:system-memory-blade:hpca:2012}) connected via RDMA (\eg,~\cite{binnig:nam:pvldb:2016, ma:asymnvm:asplos:2019, sidler:strom:eurosys:2020}) or special purpose interconnects (\eg,~\cite{novakovic:so-numa:asplos:2014, shan:legoos:osdi:2018}), or as network-attached memory controllers with more limited intelligence (\eg,~\cite{hpe:machine, aguilera:farmem:hotos:2019}).
Depending on the implementation, disaggregated memory may also be augmented with additional support for metadata operations (\eg,~\cite{tsai:atc:2020}).
Disaggregation provides separate fault domains between processing and memory, meaning that the failure of a compute node doesn't render disaggregated memory unavailable. 

\subsection{Target environment and assumptions}

In this work, we target a disaggregated memory architecture containing a high-capacity pool of memory that can be shared by compute nodes at low latency.
This memory is directly connected to the interconnect, as defined in the Gen-Z standard~\cite{GenZ}, without an explicit memory server with general-purpose processing capabilities.
Memory controllers serve non-volatile memory (NVM) on the fabric, 
which offers durability without the overheads of traditional storage devices.
Compute nodes also have local memory, which is treated as private, while the disaggregated memory is treated as shared.

Compute nodes access disaggregated memory using load and store operations, atomic memory fabric operations (\eg, compare-and-swap (\cas)) and one-sided (\eg, get and put) operations.
Disaggregated memory presents a \emph{logical address space} that is uniform across all compute nodes. Thus, any compute node can access any part of disaggregated memory in a byte-addressable manner.

Since rack-scale disaggregated NVM forms the primary persistence 
layer, it needs to be protected against memory hardware failures to avoid data loss. 
Hardware redundancy schemes have been proposed to protect both  
DRAM (\eg, ~\cite{kim:allinclusiveecc:isca:2016, zheng:raim:isca:2017}) and NVM  
(\eg, ~\cite{kateja:tvarak:isca:2020, golov:redundant-nvm:patent:2015, lesartre:fault-pmem:patent:2014}).
While these  schemes protect against memory cell or chip failures, they do not protect against failures that render a whole memory node unavailable, such as memory controller or power supply failures.
Hence, we assume a hardware redundancy scheme is used in combination with software-managed memory replication to protect disaggregated NVM. 
Using memory-side accelerators (e.g.,~\cite{daehyeok:hyperloop:sigcomm:2018}) can decentralize and minimize compute node overheads for memory replication.
Additionally, we assume that the memory interconnect incorporates sufficient component and path redundancy to prevent partitions.

\section{\MODC programming model}
\label{sec:task-model}

Programmers use resilient parallel jobs and tasks as a way to decompose and recover data processing computations. A resilient parallel application is structured as a set of restartable execution units (idempotent \emph{tasks}). 
Each task is defined as a function with an explicitly specified set of disaggregated memory-resident inputs and outputs. 
It applies a transformation on its input data to produce its output data. The output of one task may serve as an input to another task, creating an input-output \emph{data dependency}. 
Inputs and outputs are specified by names, which are resolved at runtime, thus permitting decoupled execution of producer and consumer tasks. 
A producer task can produce output data without knowing which consumer task (if any) will consume it, and a consumer task can consume data without knowing which task produced it. 
Similarly, output data may be treated as a future, allowing a consumer function to be scheduled before its inputs are fully generated; its execution is delayed until those inputs are ready (i.e., when the task(s) that produce the data are done).

\begin{figure}[t!]
\centering
\myincludegraphics{3.0in}{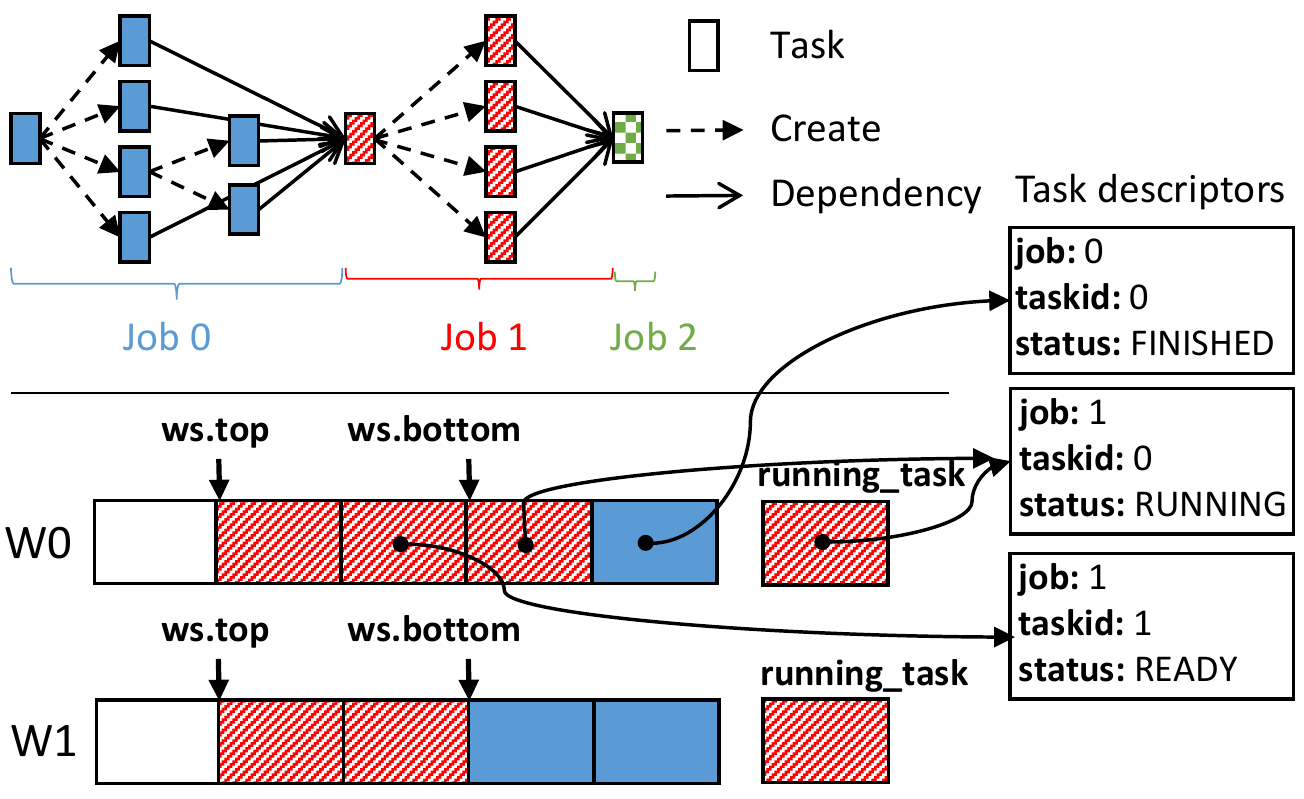}
\caption{Resilient task execution by two workers.}
\vspace{-0.2in}
\label{fig:task-parallelism}
\end{figure}

Additional \emph{control dependencies} (e.g., completion of all tasks in a set or task ordering that depends on the specific values produced rather than the existence of data) are expressed using a \emph{job} abstraction, which describes a collection of tasks. 
Control dependencies may exist between jobs, but not between tasks within a single job. An application is comprised of one or more jobs and their constituent tasks
(Figure~\ref{fig:task-parallelism}).

Expressing data dependencies with named data items rather than explicit communication between tasks means that tasks don’t need to be aware of where they (or their 
predecessors or successors) are run. 
Since tasks can be scheduled for execution on an arbitrary node, and may be restarted on a different node in case of failure, they are required to be \emph{idempotent}: given the same input, they are expected to produce the same output. 
\WORDSresponse{To enable restarting an idempotent task, the task's input data must remain available until the task successfully completes.}
Tasks are also expected to be \emph{side-effect free} in the sense that they don’t have side channels for communicating state that aren’t visible to the framework.

The programming model supports a dynamic parallelism form of autoscaling, where work can be dynamically decomposed to provide (arbitrary) levels of parallelism, depending on the amount of data to be processed. 
The programmer defines the granularity of the jobs and tasks, and thus can strike a balance on a case-by-case basis between increased parallelism, lower chance of stragglers, and reduced recovery time from smaller tasks vs. increased efficiency from larger tasks (due to lower scheduling overhead and greater execution efficiency without intra-task communication).

The application dynamically spawns tasks, and manages control dependencies while the underlying framework handles data dependencies, task scheduling, and failure recovery. 
The framework asynchronously executes each task with all-or-nothing\ignore{(at most once)} semantics, and re-schedules tasks that fail. 

\mysec{\MODC runtime}

\MODC's runtime supports the\ignore{task-based resilient} programming model using a design that avoids any centralized service that may become a performance bottleneck or single point of failure.
Space limits prevent us from describing the full runtime in detail; instead, we focus on the design principles and functionality most germane to scheduling and fault tolerance.

\mypar{Design principles} \MODC's runtime adheres to two design principles. 
\emph{First, application and runtime global state is maintained as shared data structures in disaggregated memory that are visible to all participating processes, regardless of what nodes they run on\ignore{, rather than being physically partitioned}}. 
Because all processes have direct access to global data structures, they can share data efficiently, service requests, and perform analyses for any part of a dataset, 
thus providing better load balancing and more robust performance in the face of skewed workloads.
Shared access to global data also simplifies coordination, as participants need not exchange messages to establish a common view of global state.
Processes maintain their persistent state in the shared pool, and since compute nodes and disaggregated memory fail independently, this persistent state survives the failure of any participating process or node.
If a participant dies, any other participant can pick up where the failed participant left off, simplifying failure recovery.
\emph{Second, internally, the \MODC runtime relies on lock-free synchronization for safe concurrent access to its internal data structures}.
Lock-free synchronization typically involves splitting an operation into a non-critical section activity and a short critical section, which is accomplished atomically through a fabric-supported atomic such as a compare-and-swap (\cas).
Lock-free synchronization enables some active process to make progress despite other process delays or halting failures, offering us a powerful tool to handle compute node failures.

\mypar{Design overview} \MODC's scheduler is implicit, in that a parent task creates and schedules its child tasks.
This approach supports a dynamic parallelism model, where a logical task graph is created dynamically as new tasks are scheduled. The scheduling runtime functionality is carried out by \textit{workers}, which track task dependencies, execute tasks whose dependencies are met, update scheduler metadata to report task completion, and re-execute failed tasks in the event of task, worker or compute node failures.\ignore{We assume that} The scheduling runtime\ignore{functionality} itself is fault-tolerant, written using 
lock-free programming and concurrent, shared data structures in disaggregated memory.

The \MODC framework assumes that a distributed application’s parent job is instantiated by a workload manager that is external to the \MODC runtime (e.g., \texttt{mpirun()}~\cite{Mrun}\ignore{, \texttt{shmemrun()}~\cite{SHrun}} or SLURM~\cite{SLURM}). The workload manager instantiates a per-application worker pool of processes, 
where the initial pool size is determined by the application.
The worker pool is per-application
to reduce inter-application interference. 
The pool of workers may be dynamically resized as the application executes (e.g., if more than some threshold of the original workers fails, or if the average number of ready tasks per worker exceeds or falls below a threshold). 

Each worker maintains its own task queue, which contains pointers to task descriptors for tasks that are ready to be executed (i.e., have all of their dependencies satisfied). Task queues are allocated in disaggregated memory, so that they remain accessible if a worker fails. An idle worker can steal tasks from another overloaded or failed worker. Work stealing enables dynamic load balancing and failing over a computation from a failed worker to another live worker in a decentralized manner without involving a centralized scheduler (which can become a bottleneck or single point of failure). 

\mypar{Enforcing control and data dependencies} Before a task can be added to a worker task queue, all of its control and data dependencies must be satisfied. The scheduling runtime tracks these dependencies using a collection of concurrent data structures in disaggregated memory. 
Internal scheduler metadata includes job and task descriptors containing metadata describing submitted jobs and tasks, and data structures to track progress on data and control dependencies.

To enforce control dependencies between tasks, the runtime must be able to
identify in a fault tolerant manner when all tasks of a job 
complete execution. 
We take a work-oriented approach, where a group of workers executing 
tasks of a job reach consensus that they all finished executing tasks 
in their queues through the aid of a dynamic group barrier. 
\WORDSchange{When each worker finishes executing the tasks in its own queue, it tries to steal and execute tasks from other workers' queues. When it finds no more available tasks to steal, it waits on the barrier for all other workers to finish executing their tasks or until it receives a signal to steal work from a failed worker.}

Like a traditional group barrier,  our dynamic group barrier allows members of a group to collectively reach an execution point before they proceed. 
Additionally, it allows membership to change dynamically, so that active members don't wait for failed ones.
The barrier comprises an array vector with a slot per member to indicate its participation in the barrier, two monotonically increasing sequence counters indicating membership changes and barrier releases, and a counter representing the number of members waiting on the barrier. All counters are placed contiguously in a 128-bit word so that they can be modified atomically using a 128-bit \cas. When a failed worker is detected, a participant increments the membership sequence to indicate the change. When arriving at a barrier, a member counts the number of active members and increments the waiting counter; if the active member counter matches the waiting count, then it releases the barrier.
Members spin-wait on the barrier by continuously reading the
two sequence counters for changes either in membership or barrier release. 
Upon a membership change, 
each waiting member cancels its waiting, allowing it to resume 
and help with the execution of tasks previously assigned to the failed worker.

\mypar{Executing tasks} Workers cooperatively execute tasks through work stealing, with each worker having its own task queue, 
which is structured as a 
lock-free circular-array deque~\cite{chase:wsqueue:spaa:2005}.
Each worker pulls existing tasks from and pushes newly spawned tasks to one 
end of its queue, while other workers can steal tasks from the other end (Figure~\ref{fig:task-parallelism}).
Each queue descriptor has a field that identifies
the queue's current owner worker process.
The runtime represents each task through a task descriptor that holds 
scheduling-related information, including 
the associated job, the task ID within the job, the task's execution status, and the task's serialized closure.
Each worker also maintains a slot with a pointer to
the descriptor of its last running task.
\WORDSchange{The queue, task descriptors and running slot} are allocated in disaggregated memory,
so that they remain accessible if a worker fails.

Task execution supports two stealing cases. 
First, idle workers can steal tasks from the top end of another worker's queue.
If multiple workers attempt to execute the last task in a worker's queue,
only one worker will succeed in issuing the \cas
to switch the task status from \texttt{READY} to \texttt{RUNNING}, thus guaranteeing exactly once execution. 
Second, upon a worker failure, a live worker can steal a task from  
the failed worker's running slot.
A worker that finds a running slot with a \texttt{READY} task 
can simply take over and complete the stealing procedure, as described above.
If the task's status is \texttt{RUNNING} \WORDSresponse{(i.e., the worker failed before completing the task}), then the runtime simply re-runs
the task, since tasks are considered to be idempotent.

\mypar{Detecting failed workers} The \MODC runtime detects worker process failures in a decentralized manner, where each worker may detect the failure of another.
Two global data structures provide the basic mechanism: 
a \textit{global frontier counter} and a \textit{worker vector} containing a heartbeat counter per worker to indicate the worker's notion of the current frontier. 
\WORDSchange{Each worker periodically reports its liveness status by incrementing its heartbeat counter. The global frontier counter is advanced when a sufficient number of workers reach the new value.}
Because the worker vector is stored in a designated location in 
disaggregated memory that other processes can inspect, any process can 
periodically check the heartbeat status of another process.
If an inspecting process observes that the target process' heartbeat counter is frozen for a configurable time,  
then it pronounces the target process dead \WORDSchange{and takes other corrective actions}. 

\mysec{Case study: PageRank}

\begin{figure}[t!]
\centering
\myincludegraphics{3.2in}{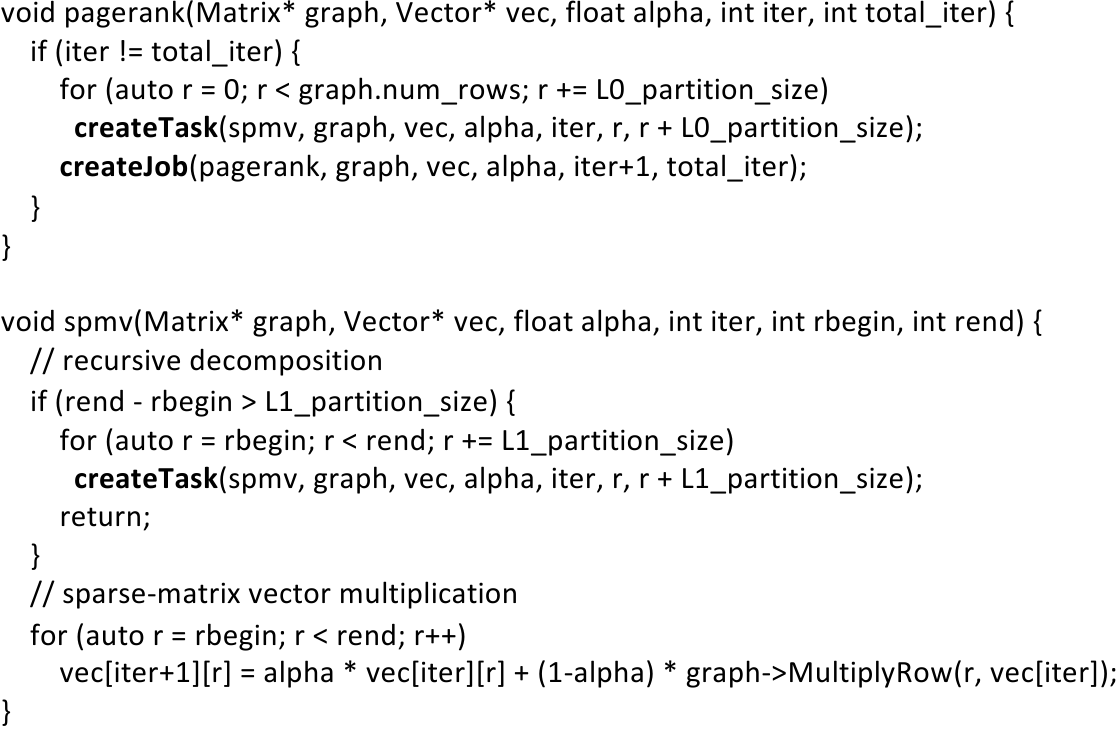}
\caption{PageRank implementation using \MODC tasks.}
\label{fig:page-rank}
\vspace{-0.2in}
\end{figure}

We demonstrate the applicability of \MODC's resilient task model to data analysis by implementing PageRank, a graph analysis algorithm that iteratively computes a rank for 
each vertex of a directed graph~\cite{page:pagerank:1999}.
Figure~\ref{fig:page-rank} shows part of a PageRank application that uses recursive decomposition to parallelize the PageRank computation in two stages\WORDSchange{, where each job corresponds to a PageRank iteration}.

First, the top-level \emph{pagerank} function performs multiple iterations, 
where each iteration $iter$ creates a fixed number of tasks; each of these tasks invokes the \emph{spmv} function. 
\emph{Pagerank} then creates a new job corresponding to the next iteration $iter+1$,
\WORDSchange{which contains a \emph{pagerank} task that will generate the tasks for iteration $iter+1$, as described above.}
Due to the control dependency between jobs, iteration $iter+1$ won’t begin until all of iteration $iter$'s tasks complete. 
Second, each \emph{spmv} task performs sparse matrix vector (SpMV) multiplication over 
its assigned part of the adjacency matrix of the graph and the current PageRank vector to generate new rank values. 
Each \emph{spmv} task may recursively (i.e., dynamically) decompose work and 
create new \emph{spmv} tasks until a target partition size is reached.

\subsection{Prototype and evaluation platform}

\WORDSresponse{
Our evaluation goal is to understand the fault tolerance % and load balancing
benefits of \MODC, as well as its overheads.
We have implemented a prototype of \MODC's resilient task programming model and runtime system, including the components described in the previous section.
Our implementation builds on several open source libraries, including lock-free data structures~\cite{RadixTree-github}, memory management~\cite{NVMM-github, librarian-lfs:2017} and fabric atomics~\cite{fam-atomics:2017}.
}

Our emulation platform for disaggregated memory is an HPE Superdome X machine running Red Hat Enterprise
Linux 7.2 with 288 cores 
and 12 TB DRAM. 
%across 16 NUMA sockets. 
Each of the 16 NUMA sockets has an E7-8890 v3 %(Haswell-EX)
CPU with 18 cores and 768 GB DRAM. 
We use Quartz~\cite{volos:quartz:middleware:2015}
to emulate disaggregated memory performance. 
Quartz emulates compute nodes by binding each\ignore{application} worker process 
to a NUMA socket, with\ignore{worker} processes evenly distributed across the\ignore{NUMA} sockets. 
It emulates disaggregated memory by binding a single instance of the \texttt{tmpfs} in-memory file system to a remote NUMA socket used exclusively for emulating disaggregated memory.
This results in 768 GB of \dm with remote latency of 400ns, as measured using a memory latency-sensitive pointer-chasing microbenchmark.
This latency is 3x slower than the local latency and on par with previous research 
proposals on rack-scale fabrics~\cite{novakovic:so-numa:asplos:2014}. 

\WORDSresponse
{Our PageRank implementations use one-dimensional row partitioning (also known as vertex partitioning): each task is responsible for a set of rows and the non-zeroes within those rows. 
Each experiment uses a $64M \times 64M$ recursive matrix (RMAT)~\cite{chakrabarti:rmat:siam:2004} graph, and results are averaged over four runs.
}

\subsection{Sensitivity to task size}
\label{sec:tasksize}

\begin{figure}[ht!]
  \centering
  \myincludegraphics{0.9\linewidth}{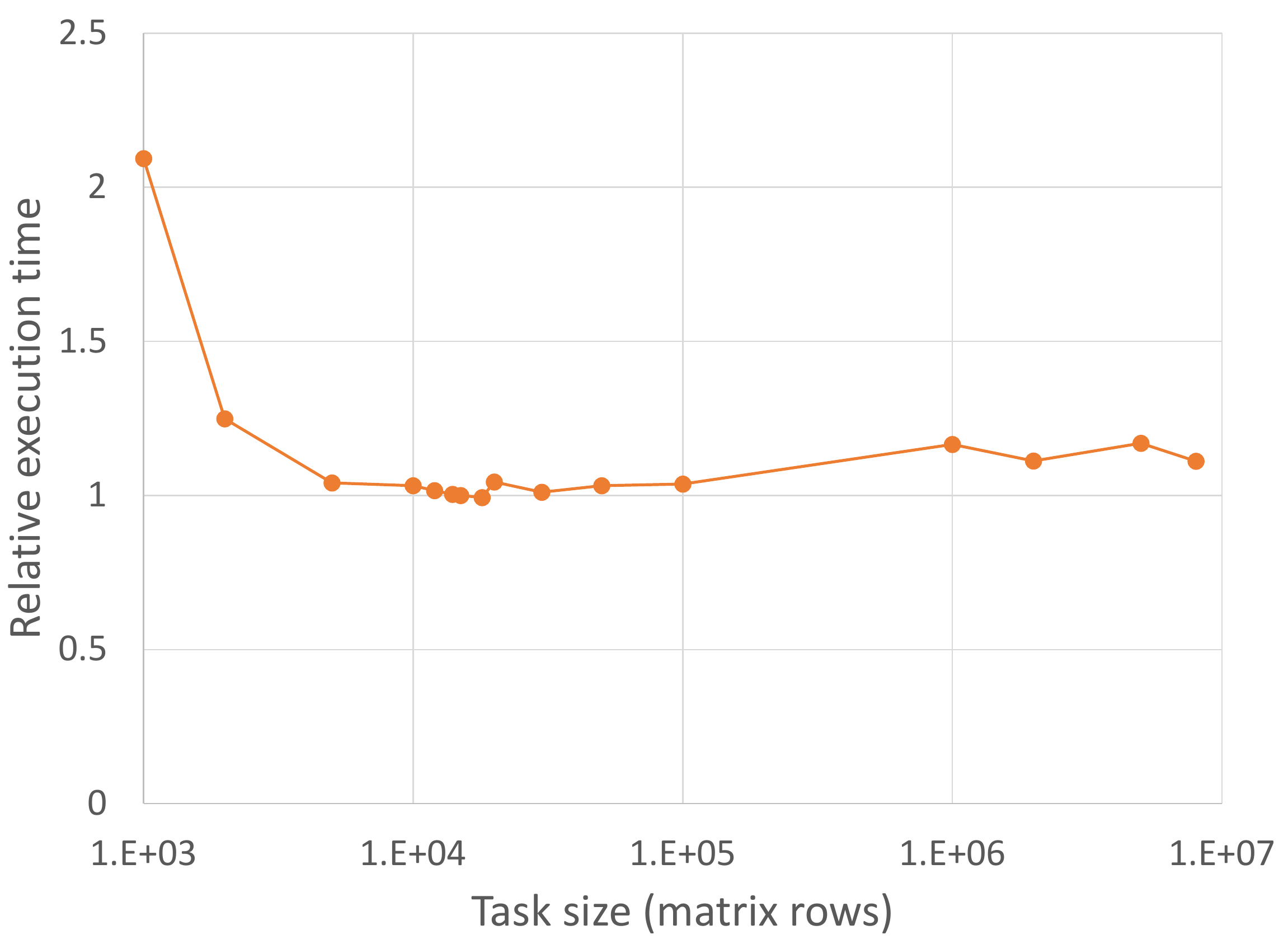}
  \vspace{-0.1in}
  \caption{\MODC PageRank sensitivity to task size}
  \label{fig:modc-tasksize}
  \vspace{-0.05in}
\end{figure}

\WORDSresponse{
We begin our analysis by exploring the sensitivity of the \MODC PageRank implementation to the task size.
Figure~\ref{fig:modc-tasksize} shows \MODC execution time relative to the best execution time (empirically determined to be 15,000 rows). This experiment uses eight workers. \MODC performs well for a wide range of task sizes: performance is within 5\% of the best for tasks of 5000 rows to 100,000 rows, and within 17\% of the best for tasks of up to 8M rows. 
We use tasks of 15,000 rows for our subsequent \MODC experiments.
}

\subsection{Fault tolerance}

We compare the resilience of the \MODC implementation of PageRank against the resilience of an MPI-based implementation with emulated checkpoints.
In both cases, fourteen workers execute PageRank’s iterative algorithm for 10 iterations.
We inject a failure by crashing one of the workers, and let each implementation recover and continue execution. 
We assume non-shrinking recovery for these experiments: a new worker is instantiated upon a worker failure, to
ensure that the number of active workers remains constant.

The MPI and \MODC implementations use the same data structures and mmap data access methods, but use distinct scheduling and recovery mechanisms.
The \MODC implementation 
dynamically decomposes the problem into jobs and tasks.
The failure is detected using \MODC's heartbeat mechanism, and a new worker is activated from a hot standby spare.
The task being executed when the worker died, along with any additional tasks in the failed worker’s queue, are stolen for execution by one of the remaining workers. 

The MPI implementation uses a bulk synchronous parallel programming model, with barriers between iterations. 
Each iteration's work is statically partitioned among the workers, with no 
work stealing. 
Partitioning assigns sets of 10,000 rows (the empirically determined best MPI set size) to workers in a round robin fashion.
For fault tolerance, the MPI implementation uses checkpoints of the dense per-iteration output vector (512 MiB) to the emulated disaggregated memory. Recovery from a failure includes the time to read the checkpoint and the time to re-run all of the iterations since the last checkpoint.
We assume a checkpoint interval of four iterations in the analysis below.

\begin{figure}[ht!]
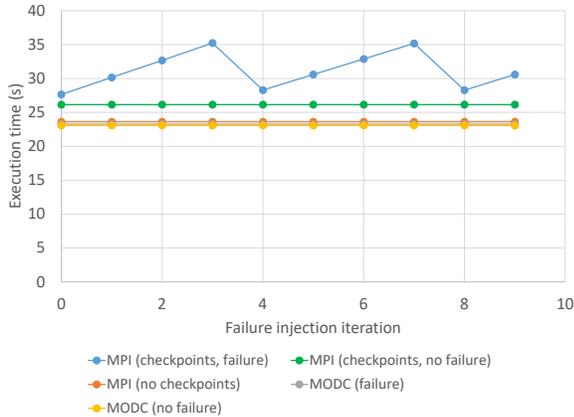

  \centering
  \myincludegraphics{0.9\linewidth}{figures/modc-mpi-ft.pdf}
  \vspace{-0.1in}
  \caption{Comparison of fault tolerance approaches}
  \label{fig:acme-mpi-ft}
  \vspace{-0.05in}
\end{figure}

Figure~\ref{fig:acme-mpi-ft} illustrates the overall execution time for\ignore{to complete} 10 PageRank iterations, with\ignore{in the face of} a failure injected at the iteration shown on the x-axis.
We measure execution with and without the failure.
With a failure injected, \MODC only needs to restart the task that was executing when the worker crashed. As a result, its overall execution time with a failure is 23.3s, less than 1\% degradation from its failure-free execution.
Execution time for MPI with checkpointing depends on how soon the crash occurs after the last checkpoint, ranging from 1.06X – 1.35X of the failure-free execution time. 
Checkpoint-recovered execution is slower than \MODC resilient execution, ranging from 19\% slower than \MODC (if the crash occurs just after the checkpoint) to as much as 51\% slower (if the crash occurs just before the next checkpoint).

\mysec{Discussion}

\WORDSchange{Although it is reasonable to ask} 
\WORDSresponse{
whether it is realistic to require programmers to structure applications as a set of idempotent tasks, this requirement is a common one among execution frameworks that target fault tolerance (e.g., CIEL~\cite{murray:ciel:nsdi:2011}, Spark~\cite{zaharia:spark:nsdi:2012} and Ray~\cite{moritz:ray:osdi:2018}), as well as recent theoretical computational models~\cite{blelloch:ppmm:spaa:2018}. We believe this model matches the needs of a large class of interesting parallel computations, including 
single program multiple data (SPMD), bulk synchronous parallel (BSP), and fork-join parallelism, as well as iterative and recursive algorithms.
}

\WORDSresponse{
The granularity of computation that makes sense as an idempotent task depends on the application's logic. As observed in Section~\ref{sec:tasksize}, very small tasks incur noticeable dependency tracking overheads, while very large tasks provide limited ability to balance load through work stealing and may increase recovery time (due to large task restart). 
\MODC successfully achieves near-optimal performance across a wide range of task sizes between these two extremes.
}

\WORDSresponse{
To address the concern of large task recovery times, we considered allowing applications to provide application-specific checkpointing, hidden from the \MODC runtime.  We found this approach inelegant, though, because it requires that the application-specific checkpoints are retained until the successful completion of the task, which requires additional runtime bookkeeping. Instead, we believe the cleanest way to avoid rework due to the restart of large tasks is for applications to use recursive decomposition to dynamically generate smaller tasks, and then let the runtime's dependency tracking handle executing and restarting the finer-grained tasks.
}

\mysec{Conclusion}
Disaggregated memory presents an opportunity to rethink how to support application resilience at rack scale.
We proposed \MODC, a framework that lets programmers write resilient applications using idempotent tasks. 
By placing the task dependency graph and work queues in disaggregated memory, \MODC's runtime can fail over computation from a failed worker to a live worker in a decentralized manner.
Our evaluation results show that 
\MODC's fine-grained resilience can outperform traditional checkpoint-based approaches in the face of failures.

\bibliographystyle{plain}
\interlinepenalty=10000
\bibliography{main}  

%%%%%%%%%%%%%%%%%%%%%%%%%%%%%%%%%%%%%%%%%%%%%%%%%%%%%%%%%%%%%%%%%%%%%%%%%%%%%%%%
\end{document}